\documentclass[aps,prd,amsmath,amssymb,reprint]{revtex4-1}

\usepackage[utf8]{inputenc}
\usepackage{amsmath}
\usepackage{color}
\usepackage{polski}
\usepackage[breaklinks]{hyperref}
\usepackage{natbib}

\begin{document}
\title{Finite Action Principle and Horava-Lifshitz Gravity: Early Universe, black-holes and wormholes.}
\author{Jan Chojnacki${}^{1}$, Jan Kwapisz${}^{1,2}$}
\date{January 2021}
\address{${}^{1}$Faculty of Physics, University of Warsaw, ul. Pasteura 5, 02-093 Warsaw, Poland \\
${}^{2}$CP3-Origins, University of Southern Denmark, Campusvej 55, DK-5230 Odense M, Denmark}

\begin{abstract}
In this work, we elaborate on the finite action in the framework of Horava-Lifshitz gravity. Assuming the Finite Action Principle we show that the beginning of the universe is flat and homogeneous. Depending on the version of the theory different cosmological scenarios are possible. Furthermore, we show that the H-L gravity action selects only the regular black-hole spacetimes, since the singular black holes possess infinite action. We also comment on the possibility of traversable wormholes in theories with higher curvature invariants. The possible cosmological solutions in Horava-Lifshitz and quadratic gravity are similar, proving that the Finite Action Principle is not model-sensitive.
\end{abstract}
\maketitle
\section{Introduction}
The path integral approach yields a powerful framework in quantum theory since it emphasises Lorentz covariance and allows for the description of non-perturbative phenomena. In the path integral, one is supposed to sum over all possible configurations of a field(s) $\Phi$ weighted by $e^{iS[\Phi]}$, where $S[\Phi]$ is the classical action of the theory. In Wick rotated path integral, where one takes $t \to i \tau$, the field configuration is weighted by $e^{-S[\Phi]}$. \\
Following this notion, the Finite Action Principle (FAP) proposes that the action should be fundamental entity instead of the field values. One can then ask which of the field configurations results in the finite action and which in the infinite one.\\
Recently, this principle has been applied to the study of black holes \cite{borissova2020blackhole}. Since it is expected that the quantum gravity should resolve the black-hole singularity problem, one may ask which of the microscopic actions remain finite for non-singular black holes and conversely interfere destructively for the singular ones. This we shall call the finite action selection principle. Only after the inclusion of higher-curvature operators, beyond the Einstein-Hilbert term, such selection principle can be satisfied \cite{borissova2020blackhole}. Furthermore in asymptotic safety, the quantum corrections to the Newtonian potential eliminate the classical-singularity \cite{Bosma:2019aiu}. \\
 On the other hand, requirement that an action of the Universe should be finite \cite{finiteaction}, is well-motivated theoretically (see also a newly proposed finite amplitudes principle \cite{Jonas:2021xkx}). This principle has a significant impact on the nature of quantum gravity and the evolution of the Universe, once the higher-curvature terms are included \cite{Lehners_2019,Barrow:2019gzc}. Following this principle, unlike for the Einstein action, in Stelle gravity \cite{Stelle:1976gc} the presence of the $R^2$ term implies homogeneous and isotropic conditions for the early universe if considering the off-shell action (for the on-shell action anisotropies are supposed to be washed out by inflation \cite{Jonas:2021xkx}). Furthermore, the highly symmetric state yields a vanishing Weyl tensor \cite{InflPenrose}, explaining the low entropy of the early universe.\\
These findings suggest that by taking into account the higher curvatures one can resolve the singularities in the early universe and the black holes. Yet, an issue with the higher-curvature theory of quantum gravity is the existence of the particles with the negative mass-squared spectrum, known as \textit{ghosts}, which makes the theory non-unitary. It is the consequence of the Ostrogradsky Theorem \cite{woodard2015theorem} and the presence of higher than second-order time derivatives in the terms beyond $R$ in the action. However this might be resolved by additional symmetry \cite{Meissner:1991zj}, giving up the micro-causality, changing the propagator prescriptions \cite{Anselmi:2018tmf,Donoghue:2019ecz} or taking into account infinitely many derivatives \cite{Barnaby:2007ve}, see also the discussion \cite{Platania:2020knd} on possible resolution in the context of asymptotic safety.\\
In this article, we explore yet another possibility, namely, we investigate Horava-Lifshitz (H-L) gravity \cite{Ho_ava_2009}, where the Lorentz Invariance (LI) is broken at the fundamental level (see \cite{Wang_2017} for a comprehensive progress report on this subject). Kinetic terms are first order in the time derivatives, while higher spatial curvature scalars regulate the UV behavior of the gravity. Furthermore, the lower-dimensional lattice studies of Causal Dynamical Triangulations (CDT) give the same Hamiltonian as H-L gravity \cite{Ambjorn:2010hu,Ambjorn:2013joa,Anderson:2011bj}.\\
In the Euclidean path integral the notion of finite action principle is quite natural, since the infinite action configurations clearly do not contribute. On the other hand, the Wick rotation is not well defined in quantum gravity in general, this become especially difficult for the theories with higher curvature terms. In our investigation we follow Horava \cite{Ho_ava_2009} and assume that Wick rotation is well defined. This can be motivated by the lack of the higher time-like curvature invariants. In the usual higher derivative constructions the existence of massive poles in the propagator makes the Wick rotation troublesome \cite{Donoghue:2019fcb}. Conversely to the latter, the new poles are massless and non-tachyonic for suitable choice of parameters \cite{Bemfica:2011xv}. Furthermore, the existence of the foliation supports that claim. For the discussion of Minkowski path integral and FAP, see \cite{borissova2020blackhole}. \\
In this article, we show that the Finite Action arguments applied to the projectable H-L gravity result in a flat, homogeneous, UV-complete, and ghost-free beginning of the universe, supporting the topological phase conjecture \cite{Agrawal:2020xek}. We also show that the Finite Action selection principle \cite{borissova2020blackhole} works for H-L gravity in the context of black holes (the action is finite for non-singular BH and conversely for the singular). Furthermore, we have found that wormholes possess a Finite Action and hence contribute to the path-integral of QG, therefore they are consistent with ER=EPR hypothesis \cite{Maldacena:2013xja}. On the other hand, the stable, traversable wormholes solutions are known only in the higher derivative gravities \cite{Duplessis_2015} (without exotic matter), so there seems to be a wormhole/non-singular BH trade-off after taking into account the Finite Action Principle. \\
\section{Horava-Lifshitz gravity}
In the Horava-Lifshitz gravity, space and time are scaled in a non-equivalent way. Diffeomorphism invariance is broken by the foliation of the 4-dimensional spacetime into 3-dimensional hypersurfaces of constant time, called leaves, making the theory power-counting renormalizable (see also the renormalization group studies of the subject \cite{DOdorico:2014tyh,DOdorico:2015pil,Barvinsky:2017kob}). The remaining symmetry respects transformations:
\begin{align}
    t\xrightarrow{}\xi_0(t),\quad x^i\xrightarrow{}\xi^i(t,x^k),
\end{align}
and is often referred to as the foliation-preserving diffeomorphism, denoted by Diff$(M,\mathcal{F})$. The diffeomorphism invariance is still present on the leaves. The four-dimensional metric may be expressed in the Arnowitt-Deser-Misner (ADM) \cite{Arnowitt_2008} variables:
\begin{align}
\label{ADM}
    (N,N^i,\,^{(3)}g_{ij}),
\end{align}
where $N,\, N^i,\,\,^{(3)}g_{ij}$ denote respectively the lapse function, shift vector, and 3-dimensional induced metric on the leaves. The theory is constructed from the following quantities:
\begin{align}
    \,^{(3)}R_{ij},\quad K_{ij},\quad a_i,\quad \,^{(3)}\nabla_i,
\end{align}
where $\,^{(3)}R_{ij}$ is the 3-dimensional Ricci curvature tensor, $\,^{(3)}\nabla_i$ is the covariant derivative constructed from the 3-dimensional metric $\,^{(3)}g_{ij}$, and $a_i:=\frac{N,_i}{N}$. Extrinsic curvature $K_{ij}$ is the only object, invariant under general spatial diffeomorphisms containing exactly one time derivative of the metric tensor $\,^{(3)}g_{ij}$:
\begin{align}
    K_{ij}=\frac{1}{2 N} \left(\frac{\partial \,^{(3)}g_{ij}}{\partial t}- \,^{(3)}\nabla_i N_j -\,^{(3)}\nabla_j N_i   \right).
\end{align}
Quantities (\ref{ADM}) are tensor/vectors with respect to Diff($M,\mathcal{F}$) possessing the following mass dimensions:
\begin{align}
    [\,^{(3)}R_{ij}]=2,\quad [K_{ij}]=3,\quad [a_i]=1,\quad [\,^{(3)}\nabla_i]=1.
\end{align}
One may use (\ref{ADM}) to construct, order by order, scalar terms appearing in the Lagrangian of the theory.  Following \cite{Wang_2017,Maier_2017} the action of the Horava gravity takes the form: 
\begin{align}
\label{ProjectableAction}
S_g = \zeta^2 \int dt dx^{3} N\sqrt{^{(3)}g} \left(\mathcal{K}-V\right),
\end{align}
where $\mathcal{K} = K_{ij}K^{ij}-\lambda K^2$ with $K=K_{ij}\,^{(3)}g^{ij} $, $\,^{(3)}g$ denotes the determinant of the 3-dimensional metric and $\zeta^2 = 1/16 \pi G$.
It may be expressed as the difference of the kinetic and potential part
$\mathcal{L}=\mathcal{K}-V$ with
$\mathcal{K}= \left(K_{ij}K^{ij}-\lambda K^2 \right)$. At the 6th order, the potential part of the lagrangian contains over 100 terms \cite{Wang_2017}. The immense number of invariants is limited by imposing further symmetries. One possible restriction for the potential comes from the projectability condition $N=N(t)$, then terms proportional to $a_i\equiv 0$ vanish. Up to the sixth order (compatible with power counting renormalizability), the potential $V$ restricted by the projectability condition is given by:
\begin{align}
\label{eq:potentialproj}
    V=2\Lambda \zeta^2-\,^{(3)}R+\frac{1}{\zeta^2} \left(g_2 \,^{(3)}R^2+g_3\,^{(3)}R^{ij}\,^{(3)}R_{ij}\right) \nonumber \\
    +\frac{1}{\zeta^4} \left(g_4{}^{(3)}R^3+g_5\,^{(3)}R\,^{(3)}R^{ij}\,^{(3)}R_{ij}+g_6\,^{(3)}R^{i}_j\,^{(3)}R^{j}_k\,^{(3)}R^{k}_i\right),\nonumber \\
     +\frac{1}{\zeta^4} \left(g_7{}^{(3)}R\nabla^2{}^{(3)}R + g_8 (\nabla_i {}^{(3)}R_{jk})(\nabla^i {}^{(3)}R^{jk}) \right),
\end{align}
where $\Lambda$ is the cosmological constant and $\alpha_{ij}$ are the coupling constants. For our purposes, we drop terms containing covariant derivatives $\,^{(3)}\nabla_i$. One should also mention that this \emph{minimal theory} \cite{Horava:2011gd} suffers from the existence of spin 0 graviton, which is unstable in the IR, see \cite{Charmousis:2009tc,Li:2009bg,Blas:2009yd,Henneaux:2009zb}. Various solutions to this problem have been proposed. One can add the additional local $U(1)$ symmetry \cite{Wang_2017,Horava:2010zj}. Then by the introduction of new fields prevents the zero-mode from propagating. On the other hand, one can drop the projectability condition $a_i=0$ and include the terms containing $a_i$ in the potential term:
\begin{align}
    V = 2\Lambda \zeta^2 -\,^{(3)}R - \beta_0 a_i a^i + \sum_{n=3}^6 \mathcal{L}_V^{(n)},
\end{align}
then for the spin-0 mode to be stable one requires $0<\beta_0<2$ \cite{Blas:2010hb,Carloni:2010ji}.\\
Note that we shall not be interested in the IR divergences stemming from boundary terms in the action and hence the famous Gibbons-Hawking-York term in context of black holes. Those divergences have nothing to do with resolution of singularities, which is an UV issue, see also discussion in \cite{borissova2020blackhole}. Nevertheless it might be useful for the reader to comment on that issue. The variational principle with Dirichlet boundary condition requires variation of the action to be zero, when we fix the boundary metric \cite{Krishnan:2016mcj}. However this is not the case for the Einstein Hilbert action and hence the famous GHY term \cite{PhysRevLett.28.1082,PhysRevD.15.2752} has to be added, which is crucial for the finiteness of the action. On the other hand for HL gravity the action possesses higher spatial derivatives. The variational principle requires $\delta \gamma_{ij}$ along with its derivatives to be zero at the spatial boundary, hence the variation is well defined without the boundary term \cite{Lara:2021jul}. We also comment further that issue in the context of cosmological spacetimes in the Sec.~\ref{Sec:Cosmo}. Furhermore absence of the boundary term has been recently proven for the mimetic Horava gravity, see \cite{Chamseddine:2019gjh,Malaeb:2020ecm}.
\section{Flatness, anisotropies and inhomogeneities in the early universe}
\label{Sec:Cosmo}
\paragraph{Flatness}
We begin our investigation of the early universe flatness by considering the FLRW metric given by the formula \cite{Kiritsis:2009sh}
\begin{align}
N\to N(t), \quad N_i \to 0, \quad {}^{(3)}g_{ij} \to a^2(t)\gamma_{ij},
\end{align}
where $\gamma_{ij}$ is a maximally symmetric constant curvature metric, with $k=+1$ for the metric on the sphere, $k=0$ for flat space time and $k=-1$ for the hyperbolic metric. We have
\begin{align}
{}^{(3)}R_{ij}= 2k\gamma_{ij}, \quad {}^{(3)}R= \frac{6k}{a(t)^2}, \quad \mathcal{K} = 3(1-3\lambda)\left( \frac{\dot{a}}{a}\right)^2,
\end{align}
and $N\sqrt{^{(3)}g} = N a^3(t)$. For $a(t)=t^s$ the kinetic part of the action gives us:
\begin{align}
\label{eq:K}
    N\sqrt{^{(3)}g}\mathcal{K}\sim t^{3s-2},
\end{align}
since $N\sqrt{^{(3)}g}\mathcal{K}\sim t^{-1}$ leads to a logarithmic divergence at $t\xrightarrow{}0$ after integrating over time, we impose that the exponent of $t$ in the integrand should be greater than $-1$. Hence for the action to remain finite as $t\to 0$, one requires $s>1/3$. In the potential part we have exemplary terms
\begin{align}
\label{eq:R}
    N\sqrt{^{(3)}g} {}^{(3)}R\sim k t^{s},\\
    \label{eq:R2}
    N\sqrt{^{(3)}g} {}^{(3)}R^2\sim k^2 t^{-s},\\
\label{eq:R3}
     N\sqrt{^{(3)}g} {}^{(3)}R^3\sim k^3 t^{-3s}.
\end{align}
For $k\neq 0$ equations (\ref{eq:K}, \ref{eq:R}, \ref{eq:R2}, \ref{eq:R3}) give rise to the following set of contradicting inequalities:
\begin{align}
    s>1/3, \quad s>-1, \quad s<1, \quad s < 1/3,
\end{align}
this shows that for $k \neq 0$ in there is no-FLRW like the beginning of the Universe for the projectable action with potential given by (\ref{eq:potentialproj}). This means, that in this version of H-L gravity, the Big Bang with power-law time dependence of the scale factor cannot be realized (similar behavior has been observed in \cite{Jonas:2021xkx} for the LI gravity). Rejecting the cubic $R^3$ terms from the potential, responsible for the contradictory inequalities, yields the action to be finite. Below we also show (on the example of Bianchi IX metric) that none of the anisotropic non-flat solutions are allowed in the action with terms cubic in Ricci curvature.
\paragraph{Anisotropies}
\label{par:Anisotropies}
We consider Bianchi IX metric as a representative model of non-flat anisotropic spacetimes (in this paragraph $k=1$): 
\begin{align}
ds_{IX}^2=-N^2 dt^2+h_{ij}\omega^i \omega^j, 
\end{align}
where $h_{ij}=diag\left(M^2, Q^2, R^2 \right)$ and $M,Q,R$ are functions of the time only. The connection is 
\begin{align}
    d\omega^a=\Gamma^a_c\wedge \omega^c=\Gamma^a_{cb}\omega^b \wedge \omega^c .
\end{align}
The Bianchi IX one forms satisfy:
\begin{align}
\label{eq:BianchiIXforms}
    d\omega ^a=\frac{1}{2}\epsilon^{abc}\omega^b \wedge \omega^c.
\end{align}
Hence $\Gamma^a_{bc}=-\frac{1}{2}\epsilon^{abc}$. The usual closed FRLW universe is obtained when $R(t)=M(t)=Q(t)=\frac{a(t)}{2}$, where $a(t)$ is the scale factor. The explicit form of the curvature invariants was calculated in \cite{Maier_2017}:
\begin{align}
    \,^{(3)}R&=\frac{-1}{2M^2Q^2R^2}\Big(M^4+Q^4+R^4-\left(R^2-Q^2 \right)^2\nonumber\\&-\left(R^2-M^2 \right)^2-\left(M^2-Q^2 \right)^2 \Big),
\end{align}
\begin{align}
    \,^{(3)}R^{i}_j\,^{(3)}R^{j}_i&=\frac{1}{4(M Q R)^4}  \Big[3M^8-4M^6\left(Q^2+R^2\right)\nonumber\\&-4M^2 \left(Q^2-R^2\right)^2\left(Q^2+R^2\right)\nonumber\\&+2M^4\left(Q^2+R^2\right)^2+\left(Q^2-R^2\right)^2\left(3Q^4\right)\nonumber\\&+ \left(Q^2-R^2\right)^2\left(2Q^2R^2+3R^4 \right) \Big],
\end{align}
\begin{align}
    \,^{(3)}R^{i}_j\,^{(3)}R^{j}_k\,^{(3)}R^{k}_i&=\frac{1}{8(MQR)^6}\Big(\left[ (M^2-Q^2)^2-R^4\right]^3 \nonumber\\&+ \left[ (M^2-R^2)^2-Q^4\right]^3 \nonumber\\&+\left[ (Q^2-R^2)^2-M^4\right]^3\Big).
\end{align}
The kinetic and the potential part are respectively:
 \begin{align}
   N\sqrt{^{(3)}g}  \mathcal{K}&=\frac{MQR}{N}\Big[\left(1-\lambda\right)\left(\frac{\dot M^2}{M^2}+\frac{\dot Q^2}{Q^2}+\frac{\dot R^2}{R^2} \right)\nonumber\\&-2 \lambda \left( \frac{\dot M \dot Q}{MQ}+\frac{\dot Q \dot R}{QR}+\frac{\dot M \dot R}{MR}\right) \Big],\\
    N\sqrt{^{(3)}g}  V&=-N\left(MQR\right) V.
\end{align}
For the Bianchi IX metric we use the following ansatz:
\begin{align}
    M(t)\sim t^m,\quad Q(t)\sim t^q,\quad R(t)\sim t^r.
\end{align}
With such solutions, the kinetic term is proportional to
\begin{align}
    N\sqrt{^{(3)}g}\mathcal{K}\sim t^{m+q+r-2}.
\end{align}
This results in an inequality:
\begin{align}
\label{KineticCondition}
    m+q+r>1.
\end{align}
Similar reasoning is applied to all of the curvature scalars in the potential. Ricci scalar terms lead to conditions:
\begin{align}
    3m-q-r>-1,\quad 3q-m-r>-1,\quad 3r-m-q>-1,\\ r+q-m>1,\quad r+m-q>-1,\quad m+q-r>-1.
\end{align}
Quadratic terms are numerous and we provide explicit conditions only for the $R_{ij}R^{ij}$ terms:
\begin{align}
&    5m-3q-3r>-1,\quad 3m-q-3r>-1,\quad 3m-3q-r>-1,\nonumber\\
 &   3r-m-3q>-1,\quad r-m-q>-1,\quad q-m-r>-1,\\
  &  3q-m-3r>-1,\quad m+q-3r>-1,\quad m-q-r>-1,\\
   & m+r-3q>-1,\quad 5q-3m-3r>-1,\quad 3q-3-r>-1,\\
    &q+r-3m>-1,\quad 3r-q-3m>-1,\quad -3m-3q+5r>-1.
\end{align}
It is tedious to algebraically verify that the above set of conditions is not contradictory. A geometrical interpretation brings more light to the problem: each of the inequalities corresponds to a half of the $\mathbb{R}^3$ space in $(q,m,r)$ coordinates. The common subspace restricted by a pair of such inequalities vanishes if the planes corresponding to the boundary of the half-spaces are parallel. This is easily verified by considering the vector normal to each plane. For example the boundary plane obtained from inequality $m+q+r>1$ is $m+q+r=1$ and the normal vector $(1,1,1)$. If two such normal vectors are parallel (bearing in mind the correct inequality direction), the half-spaces will be separate.\\
The kinetic part and scalars up to quadratic order in curvature do not lead to contradictory conditions. However, including the $R^3$ term we have:
\begin{align}
    N\sqrt{^{(3)}g}\,^{(3)}R^3 \ni \frac{N}{MQR}\sim t^{-m-q-r}\implies m+q+r<1.
\end{align}
This is in clear contradiction with (\ref{KineticCondition}).
This means, that also Bianchi IX anisotropic spacetime leads to infinite action. Notice, that taking the isotropic limit $m=q=r$ also leads to infinite action, as discussed in the previous paragraph. \\ 
There are two ways of dealing with this - leaving the model unchanged and considering other cosmological solutions, such as oscillating universe with bounded $a(t)\in(a_{min},a_{max})$, see for example \cite{Fukushima:2018xgv}. 
On the other hand,  this might be understood as indication that the flatness problem is resolved via the Finite Action Principle without the need of inlfation. In  particular for flat, anisotropic Bianchi I spacetime, all of the spatial invariants vanishes leaving us with the kinetic term condition
\begin{align}
    m+q+r>1.
\end{align}
Therefore, the action is finite for both isotropic and anisotropic flat beginning of the universe in contradistinction to the Lorentz covariant $R^2$ off-shell action \cite{Lehners_2019}, yet during the evolution of the universe, the anisotropies might vanish dynamically \cite{Jonas:2021xkx} if spacetime is accelerating. Furthermore, this is generically true for $\Lambda>0$, see the dynamical systems studies of the matter \cite{Misonoh:2011mn}. In our analysis we have neglected the boundary terms in the action. On non-flat FLRW spacetime they could lead only to further contradictory conditions. On the other hand for $k=0$ FLRW solution HL gravity those terms will play no role, as it was investigated in \cite{Jonas:2021xkx}. 
\paragraph{Inhomogeneities}
Unlike the anisotropies, the finiteness of the action suppresses the inhomogeneities already at the second-order of the spatial Ricci scalar curvature.
Investigation of the inhomogeneities concerns following isotropic metric tensor:
\begin{align}
    ds^2=-dt^2+\frac{A'^2}{F^2}dr^2+A^2\left(d\theta^2+\sin^2\theta d\phi^2\right),
\end{align}
where $A=A(t,r)$, $F=F(r)$ and $A'=\partial_r A$. The homogeneous FRLW metric is retrieved, when $F\xrightarrow{}1$. The resulting Ricci scalar and Ricci scalar squared contribution to the action are:
\begin{align}
\sqrt{\,^{(3)}g}\, ^{(3)}R\sim 2AF'+\frac{A'\left(-1+F\right)^2}{F} ,
\end{align}
\begin{align}
 \sqrt{\,^{(3)}g}\, ^{(3)}R^2\sim\frac{2AFF'+A'\left(F^2-1 \right) }{A^2A'F}.   
\end{align}
Again, we suppose that each term should be convergent as $t\xrightarrow{}0$.
By the ansatz $A(t)\sim t^s$, inequalities stemming from $\,^{(3)}R$ and $\,^{(3)}R^2$ are contradictory. This means, that $F(r)\xrightarrow{}1$, hence the metric of the early universe was homogeneous.
\section{Black holes and wormholes}
In this section, we show that H-L gravity satisfies the Finite Action selection principle for the microscopic action of quantum gravity \cite{borissova2020blackhole}. We study both the solutions of H-L gravity and the known, off-shell BH spacetimes, due to the fact that there are no known regular BH solutions in the HL gravity. Keep in mind that a metric does not need to be a solution to the equations of motion to enter the path integral. We require singular black-hole metrics to interfere destructively, while the regular ones with finite action contribute to the probability amplitudes. We broaden this analysis by studying the wormhole solutions. 
\subsection{Singular black holes} 
Singularities may be categorised \cite{Ellis:1977pj} in the three main groups: \textit{scalar, non-scalar} and \textit{coordinate singularities}. Scalar singularities are the ones for which (some of) the curvature invariants, like Kretschmann scalar, become divergent and hence they are the object of interest in our considerations. Non-scalar singularities appear in physical quantities such as the tidal forces. Finally, the coordinate singularities appear in the metric tensor, however, one may get rid of the divergence with a proper coordinate transformation. Yet, coordinate singularities of General Relativity may become scalar singularities in the Horava-Lifshitz gravity \cite{Cai_2010}. It is due to the fact that the spacetime diffeomorphism of GR is a broader symmetry than the foliation-preserving diffeomorphism of H-L gravity. As an example, consider the Schwarzschild metric:
\begin{align}
\label{Schw}
    ds^2=-\left(1-\frac{2m}{r} \right)dt^2+\left(1-\frac{2m}{r}\right)^{-1}dr^2+r^2d\Omega^2,
\end{align}
where $d\Omega^2=r^2(d\theta^2+\sin^2{\theta}\,d\phi^2)$.
The singular points are $r=0$ and $r=r_s=2m$. In GR the singular point $r_s=2m$ may be removed by the transformation
\begin{align}
\label{PGtransformation}
    dt_{PG}=dt+\frac{\sqrt{2mr}}{r-2m}dr.
\end{align}
The resulting Painleve-Gullstrand metric is:
\begin{align}
\label{PG}
    ds^2=-dt^2_{PG}+\left(dr-\sqrt{\frac{2m}{r}}dt_{PG} \right)^2+r^2 d\Omega^2.
\end{align}
For more details see e.g. \cite{Martel_2001}.
In GR metric tensors (\ref{Schw}) and (\ref{PG}) describe the same spacetime with singularity at $r=0$. Notice, however, that the coordinate transformation (\ref{PGtransformation}) does not preserve the spacetime foliation, breaking the projectability condition. Hence, in the framework of H-L gravity, metric tensors (\ref{Schw}) and (\ref{PG}) describe distinct spacetimes. Moreover, as we will show, Schwarzschild's metric singularity at $r=r_s$ becomes a spacetime singularity. Hence due to the unique nature of the foliation-preserving diffeomorphism, investigating the singularities in H-L gravity is a delicate matter.\\ 
We consider three representative solutions \cite{Cai_2010}: (anti-) de Sitter Schwarzschild, which is the simplest spacetime with the black hole and the cosmological horizon, Kerr spacetime (see also the rotating H-L solution \cite{Wang:2012nv}) and the H-L solution found by Lu, Mei and Pope (LMP) \cite{L__2009}. In this section, we discuss the (anti-) de-Sitter Schwarzschild as the clearest example. The other metrics have similar features and we explore them in the Appendix. In the following, all of the curvature scalars are three-dimensional, unless stated otherwise. 
\paragraph{(Anti-) de Sitter Schwarzschild solution}
The general static ADM metric with projectability condition takes the form:
\begin{align}
\label{ADMgauge}
    ds^2=-dt^2+e^{2\nu}(dr+e^{\mu-\nu}dt)^2+r^2d\Omega^2,
\end{align}
where $\mu=\mu(r)$, $\nu=\nu(r)$. (Anti-) de Sitter Schwarzschild solutions are be obtained for:
$\mu=\frac{1}{2}\ln{\left(\frac{M}{r}+\frac{\Lambda}{3}r^2 \right) }$, $\nu=0.
$ The resulting kinetic terms and Ricci scalar are:
\begin{align}
\label{ADMSch}
   {}^{(3)}R&=0,\nonumber\\
    K&=\left(\frac{3M+\Lambda r^3}{12 r^3} \right)^\frac{1}{2}\left(\frac{4}{r}-\frac{3M-2\Lambda r^3}{3M+\Lambda r^3} \right),\nonumber\\
    K_{ij}K^{ij}&=\frac{3M+\Lambda r^3}{12 r^3}\left[8+\left(\frac{3M-2\Lambda r^3}{3M+\Lambda r^3}  \right)^2 \right].
\end{align}
The kinetic part is divergent at $r=0$ and $r=\left(\frac{3M}{|\Lambda|}\right)^\frac{1}{3}$ for the negative cosmological constant. We investigate the finiteness of the function:
\begin{align}
\label{SSFunction}
S_s(r_{UV},r_{IR}):=\int_{r_{UV}}^{r_{IR}}dr\,N\sqrt{g}\left(K_{ij}K^{ij}-\lambda K^2+ {}^{(3)}R\right),
\end{align}
which is a part of the action qualitatively describing the singularities. The $r_{IR}$ is chosen so that the volume integral is finite, hence we do not consider singularities stemming from the IR behaviour (large distances) of the spacetime boundary and time integration. In particular The $r_{UV}$ is the minimal radius, which we take $r_{UV} \to 0$. For the scalars (\ref{ADMSch}) and value of $\lambda \neq 1$, the function $S_s(r_{UV},r_{IR})$ is divergent at the expected points $r_s=r_{UV}=0$ and $r_s=\left(\frac{3M}{|\Lambda|}\right)^{\frac{1}{3}}$. However, for $\lambda=1$, which is the value required for low energy Einstein-Hilbert approximation, the terms divergent at $r_s=\left(\frac{3M}{|\Lambda|}\right)^{\frac{1}{3}}$ remain finite as one could expect, since $r_s$ corresponds to the cosmological horizon. Explicitly we have:
\begin{align}
S_s(r_{UV},r_{IR})&=\frac{2}{9}\Lambda r^3_{UV}-8\Lambda r^2_{UV}+\left( \frac{M}{4}-16\Lambda\right)\ln{r_{UV}}\nonumber\\&-\frac{24M}{ r_{UV} }+\frac{24}{r^2_{UV}}+ \textrm{IR terms}.
\end{align}
Here, only the spatial Ricci scalar is necessary for the singular solution to be suppressed in the gravitational path integral.\\  
As mentioned previously, different gauges of the same spacetime in GR, correspond to distinct spacetimes in H-L gravity. Hence, we consider the (anti-) de Sitter Schwarzschild metric in the orthogonal gauge, which is not a solution to the projectable H-L theory, in contrast to the previous case: 
\begin{align}
\label{orthogonalgauge}
    ds^2=-e^{2\Psi(r)}d\tau^2+e^{2\Phi(r)}dr^2+r^2 d\Omega^2,
\end{align}
here
\begin{align}
    \Psi(r)=-\Phi(r)=\frac{1}{2}\ln\left(1-\frac{2 M}{r}+\frac{1}{3}\Lambda r^2 \right).
\end{align}
In the orthogonal gauge, the components of the metric tensor do not depend on the time coordinate, hence the kinetic part vanishes $K_{ij}=0$. One finds, that the Ricci scalar is constant $ {}^{(3)}R=-2\Lambda$. However, the higher-order curvature terms (see Appendix for the general form) are divergent at the origin:
\begin{align}
     {}^{(3)}R_{ij} {}^{(3)}R^{ij}&=\frac{4 \Lambda ^2}{3}+\frac{6 M^2}{r^6},\nonumber\\
     {}^{(3)}R^i_{j} {}^{(3)}R^j_k {}^{(3)}R^k_i&=-\frac{8 \Lambda ^3}{9}-\frac{12 \Lambda  M^2}{r^6}-\frac{6 M^3}{r^9},
\end{align}
yielding an infinite action and suppressing the singularity. The same conclusions can be drawn for Kerr spacetime and singular Lu-Mei-Pope metric, derived in the context of Horava gravity, see Appendix.
\\
\paragraph{Regular black holes}
Due to observation's of the binary black holes mergers \cite{PhysRevLett.116.061102} and the Event Horizon Telescope observations \cite{Akiyama:2019cqa,Akiyama:2019eap} the structure of BH can be investigated on an unprecedented scale \cite{Dymnikova:2019vuz}. Furthermore, due to the expectation that the quantum gravity shall resolve the BH singularity issue, the regular black holes have been of interest recently, for discussions in various quantum gravity approaches \cite{Bonanno:2000ep,Ashtekar:2005qt,Modesto:2005zm,Bonanno:2006eu,Falls:2010he,Held:2019xde,Platania:2019kyx,Faraoni:2020stz} (see for more model independent viewpoint \cite{Dymnikova:1992ux,Hayward:2005gi,Bambi:2013ufa,Frolov:2016pav}). Following \cite{borissova2020blackhole}, we shall discuss Hayward metric \cite{Hayward:2005gi} (Dymnikova spacetime \cite{Dymnikova:1992ux} is discussed in the Appendix).  The Hayward metric is an example of the regular black hole solution in GR:
\begin{align}
\label{Heyward}
    ds^2&=-f(r)dt^2+f(r)^{-1}dr^2+r^2d\Omega^2,\nonumber\\
    f(r)&=1-\frac{2 M r^2}{\left(r^3+2g^3\right)},
\end{align}
where $g$ is an arbitrary positive parameter. The metric is non-singular in $r\xrightarrow{}0$. It is not a solution to H-L theory, however, we consider it as an off-shell metric present in the path integral.\\
The kinetic tensor vanishes $K_{ij}=0$, while the Ricci scalar and the second-order curvature scalars are regular:
\begin{align}
    {}^{(3)}R&=\frac{24 g^3 GM}{(2g^3+r^3)^2},\nonumber\\
    {}^{(3)}R_{ij}{}^{(3)}R^{ij}&=\frac{6 M^2 \left(32 g^6+r^6\right)}{\left(2 g^3+r^3\right)^4},
\end{align}
leading to finite action. A similar conclusion arises in the case of Dymnikova spacetime, see Appendix. These two regular solutions to GR are also regular in the off-shell H-L theory.
\subsection{Wormholes}
Here, we take the first step in the direction of the investigations of the consequences of the Finite Action Principle in the context of wormholes (WH). The wormholes may be characterized in two classes: traversable and non-traversable. The traversable WH, colloquially speaking, are such that one can go through it to the other side, see \cite{Morris:1988cz} for specific conditions.
The pioneering Einstein-Rosen bridge has been found originally as a non-static, non-traversable solution to GR. The traversable solutions are unstable, however, they might be stabilized by an exotic matter or inclusion of the higher curvature scalar gravity \cite{Duplessis_2015}. This is important in the context of finite action since usually the divergences of black holes do appear in the curvature squared terms. Hence, due to the inclusion of the higher-order terms in the actions, the traversable wormholes are solutions to the equations of motions without the exotic matter. The exemplary wormhole spacetimes investigated here are the Einstein-Rosen bridge proposed in \cite{PhysRev.48.73}, the Morris-Thorne (MT) wormhole \cite{Morris:1988cz}, the traversable exponential metric wormhole \cite{Boonserm_2018} and the wormhole solution discussed in the H-L gravity \cite{Botta_Cantcheff_2010}. All of them have a finite action. Here, we shall discuss the exponential metric WH. The conclusions for the other possible wormholes are similar and we discuss them in the Appendix. For the exponential metric WH, the line element is given by:
\begin{align}
    ds^2=-e^{-\frac{2M}{r}}dt^2+e^{\frac{2M}{r}}\left(dr^2+r^2 d\Omega^2 \right).
\end{align}
 This spacetime consists two regions: ``our universe'' with $r>M$ and the ``other universe" with $r<M$. $r=M$ corresponds to the wormhole's throat. The spacial volume of the ``other universe" is infinite when $r\xrightarrow{}0$. Such volume divergence is irrelevant to our discussion since it describes large distances in the "other universe". Hence, we further consider only $r\geq M$. 
The resulting Ricci and Kretschmann scalars calculated in \cite{Boonserm_2018} and the measure are non-singular everywhere:
\begin{align}
    R&=-\frac{2M^2}{r^4} e^{\frac{-2M}{r}},\nonumber\\
    R_{\mu\nu\sigma\rho}R^{\mu\nu\sigma\rho}&=\frac{4M^2 (12r^2-16Mr+7M^2)}{r^8}e^{-4\frac{M}{r}},
\end{align}
resulting in the finite action for the Stelle gravity. Similarly for the H-L gravity:
\begin{align}
    {}^{(3)}R=R,\quad K^2=K_{ij}K^{ij}=0,\nonumber\\
    {}^{(3)}R_{ij} {}^{(3)}R^{ij}=\frac{2 M^2 (M^2-2 M r+3 r^2)}{r^8}e^{-\frac{4 M}{r}}.
\end{align}
\section{Conclusions and discussion}
 The Finite Action Principle is a powerful tool to study quantum gravity theories and also QFTs in general. In particular, we have shown that it can be invoked to explain the flatness and homogeneity of the early universe and can possible be a dynamic resolution of the singularities  problem of black holes in the context of Horava-Lifszyc gravity.\\
The conditions stemming from the Finite Action Principle justify the Topological Phase hypothesis without the need for conversion of the degrees of freedom in the early universe, which is assumed to take place in \cite{Agrawal:2020xek}. Furthermore, the anisotropic scaling of Horava gravity admits only flat solutions for the cosmological metrics, see also the discussion on the instanton ``no-boundary"-like solution \cite{Bramberger:2017tid}. Moreover, the amplitude of the cosmological perturbations are  scaling as: 
$\delta \Phi = H^{\frac{3-z}{2z}},$ hence at $z=3$ they are almost scale-invariant \cite{Mukohyama:2009gg}. Finally, the Weyl anomalies structure in H-L gravity does not lead to strong non-local effects during the radiation domination epoch \cite{Godazgar:2016swl,Meissner:2016onk}. This stems from the fact that these anomalies are of second order in derivatives in the flat spacetimes \cite{Griffin:2012qx,Arav:2014goa,Arav:2016akx,Arav:2016xjc}, hence they are harmless and allow to avoid the vanishing of conformal anomaly criteria \cite{Meissner:2017qwm,Meissner:2018hmx}. In particular, it would be interesting to see whether the anisotropic Weyl anomalies can also give departure from scale invariance, as it is discussed in \cite{Agrawal:2020xek}. Yet we leave that for further investigation to be performed elsewhere.
Combined with earlier results, our investigation backs up fully the topological phase conjecture, hence making inflation redundant. Furthermore, it seems that this is in line with the swampland conjectures and the newly proposed finite-amplitude principle 
\cite{Jonas:2021xkx}, making the asymptotically safe quantum gravity to pick initial conditions such that inflation ceases to be eternal \cite{Chojnacki:2021fag}, see also \cite{Rudelius:2019cfh,Rudelius:2021oaz}.\\
From the point of view of finite action selection principle \cite{borissova2020blackhole} they are equally good theories, resolving the black holes singularities, assuming that ghost issue is resolved in the latter case. Yet none of the regular B-H solutions have been found in the context of H-L gravity \cite{AW}. Hence it is a strong suggestion that the wormholes may appear in the UV regime of H-L gravity and can serve as a ``cure'' for singularities \cite{Lobo:2016zle,Olmo:2016hey,Bambi:2015zch}.\\
In the case of wormholes, both traversable and non-traversable wormholes are on equal footing in the case of the Finite Action Principle. However, this principle suggests that there is a trade-off between the resolution of black-hole singularities and the appearance of wormhole spacetimes due to higher curvature invariants. The wormhole solutions will remain in both the LI and H-L path integrals. The higher-order curvature scalars, generically present in the quantum gravity, stabilize the wormhole solutions without the need for an exotic matter.\\
Finally, one should mention that there are many experiments to test the Lorentz Invariance Violations (LIV) in the gravitational sector coming from gravitational waves observations \cite{Yunes:2016jcc,Gumrukcuoglu:2017ijh,Abbott:2018lct,Ramos:2018oku,Mewes:2019dhj,Frusciante:2020gkx}, which could in principle validate Horava's proposal, yet we know much more about the LIV in the matter sector (see for example \cite{Tasson:2014dfa,Frusciante:2015maa}). Since these two can be related \cite{Eichhorn:2019ybe}, one can speculate that H-L gravity can be tested in the nearby future.\\
\subsection{Acknowledgements} We thank J. N. Borrisova, A. Eichhorn, L. Giani, K. A. Meissner, A. G. A. Pithis and A. Wang for inspiring discussions and careful reading of the manuscript. J.H.K. was supported by the Polish National Science Centre grant 2018/29/N/ST2/01743. J.H.K. would like to thank the CP3-Origins for the extended hospitality during this work. J.H.K acknowledges the NAWA Iwanowska scholarship PPN/IWA/2019/1/00048.

\appendix
\setcounter{section}{0}
\setcounter{equation}{0}
\renewcommand\thesection{\Alph{section}}
\section{Appendix: Further black-holes and wormholes}

Here we present further examples of interesting black-hole and wormhole spacetimes in the context of the Finite Action Principle. We find the restrictions on the LMP solutions necessary to resolve the singularity at the origin. Similarly, the spatial Ricci scalar of Kerr's spacetime yields infinite action. We further give three examples of wormholes with finite action: Einstein-Rosen bridge, Morris-Thorne wormhole, and a spatially symmetric and traversable wormhole solution to H-L gravity.
\subsection{Black-holes}
\renewcommand{\theequation}{A-\arabic{equation}}
\paragraph{LPM black hole}
The popular LMP \cite{L__2009} metric is not a solution to the vacuum H-L equations. However, the second class of the LMP solutions written in the ADM frame with projectability condition satisfy the field equations of H-L gravity coupled to anisotropic fluid with heat flow, see \cite{Cai_2010}. The LMP solutions were found in the orthogonal gauge (\ref{orthogonalgauge}), without the projectability. There are two types of solutions. Class A solutions are:
\begin{align}
\label{ClassA}
    \Phi=-\frac{1}{2}\ln (1+x^2),\quad \Psi=\Psi(r).
\end{align}
Class B solutions consist of:
\begin{align}
    \label{ClassB}
    \Phi&=-\frac{1}{2}\ln \left(1+x^2-\alpha x^{\alpha_{\pm}}\right),\nonumber\\
    \Psi&=-\beta_{\pm}\ln{x}+\frac{1}{2}\ln \left(1+x^2-\alpha x^{\alpha_{\pm}}\right),
\end{align}
where $x=\sqrt{|\Lambda_W|}r$, $\Lambda=\frac{3}{2}\Lambda_W$, $\alpha$ is an arbitrary real constant, and $\alpha_{\pm}$ and $\beta_{\pm}=2\alpha_{\pm}-1$ are parameters depending on $\lambda$. Their explicit form may be found in \cite{Cai_2010}. The LPM solutions (\ref{ClassA}) and (\ref{ClassB}), have vanishing kinetic tensor $K_{ij}=0$, while the Ricci scalar and the integral measure are given respectively by:
\begin{align}
    {}^{(3)}R=\frac{2}{r^2}\left( \alpha (1+\alpha_{\pm})x^{\alpha_{\pm}}-3x^2\right),\quad N\sqrt{g}=r^2x^{-\beta_{\pm}}.
\end{align}
The $S_s$ function (\ref{SSFunction}) stands:
\begin{align}
    S_s(x_{UV},x_{IR})&=-\frac{2}{\sqrt{\Lambda_W}}\int_{x_{UV}/\sqrt{\Lambda_W}}^{x_{IR}/\sqrt{\Lambda_W}}dx\,(\alpha(1+\alpha_{\pm})x^{1-\alpha_{\pm}}\nonumber\\&-3x^{3-2\alpha_{\pm}}),
\end{align}
Where $x_{UV}=\sqrt{\Lambda_W}r_{UV}$ and $x_{IR}=\sqrt{\Lambda_W}r_{IR}$.
The necessary condition for the spatial Ricci scalar to be finite is $2>\alpha_{\pm}$.
We proceed in the ADM gauge, which describes an independent theory in the H-L gravity. Then, the Class A solution is given by:
\begin{align}
    \mu=-\infty,\quad \nu=-\frac{1}{2}\ln\left(1-\Lambda_W r^2 \right)
\end{align}
applied to (\ref{ADMgauge}), we get $ {}^{(3)}R=6\Lambda_W$. The $S_s$ function is given by:
\begin{align}
    S_s(r_{UV},r_{IR})=-\int_{r_{UV}}^{r_{IR}}\frac{6\Lambda_W r^2}{\sqrt{|1-\Lambda_W r^2|}}.
\end{align}
The exact form of $S_s(r_{UV},r_{IR})$ depends on the sign of the scaled cosmological constant $\Lambda_W$, nevertheless, it is always finite, when $r_{UV}\xrightarrow[]{}0$. Indeed, for the negative $\Lambda_W<0$:
\begin{align}
     S_s(r_{UV},r_{IR})&=-\frac{3}{\sqrt{-\Lambda_W}}\textrm{arcsinh}\left(\sqrt{-\Lambda_W}r_{UV}\right)\nonumber\\&-3r_{UV}\sqrt{-\Lambda_W r^2_{UV}+1}.
\end{align}
Positive cosmological constant splits the space in two regions. When $r>\frac{1}{\sqrt{\Lambda_W}}$ we get:
\begin{align}
    S_s(r_{UV},r_{IR})&=\frac{3}{\sqrt{\Lambda_W}}\textrm{arctanh}\left( \frac{\sqrt{\Lambda_W}r_{UV}}{\sqrt{\Lambda_W r^2_{UV}-1}}\right)\nonumber\\&+3r_{UV}\sqrt{\Lambda_W r^2_{UV}-1},
\end{align}
for a small, positive cosmological constant, above result is irrelevant for our discussion, since it would describe large scales.
When $r<\frac{1}{\sqrt{\Lambda_W}}$:
\begin{align}
    S_s(r_{UV},r_{IR})&=-\frac{3}{\sqrt{\Lambda_W}}\textrm{arcsin}\left(\sqrt{\Lambda_W}r_{UV}\right)\nonumber\\&+3r_{UV}\sqrt{\Lambda_W r^2_{UV}-1}.
\end{align}
The class B solution singularity at the origin, appearing when $ 2\leq \alpha_{+}$ is suppressed by the Finite Action Principle. Class A solutions are finite and contribute to the path integral, if the cosmological constant is negative or small and positive when $r_{UV}\xrightarrow{}\frac{1}{\sqrt{\Lambda_{W}}}$.
\paragraph{Kerr spacetime}
Kerr spacetime corresponds to an axially symmetric, rotating black hole with mass $M$ and angular momentum $J$. It is a solution to the Einstein Equations in GR, however, it has been shown order by order in the parameter $a=J/M$, that it is not a solution to the H-L field equations \cite{Lee_2012}. Yet it can still enter the path integral as an off-shell metric. The line element in the Boyer-Lindquist coordinates is given by:
\begin{align}
    ds^2=-\frac{\rho^2 \Delta_r}{\Sigma^2}dt^2+\frac{\rho^2}{\Delta_r}dr^2+\rho^2 d\theta^2+\frac{\Sigma^2 \sin^2{\theta}}{\rho^2}(d\phi-\xi dt^2)^2,
\end{align}
where
\begin{align}
    \rho^2&=r^2+a^2 \cos^2{\theta},\nonumber\\
    \Delta_r&=r^2+a^2-2Mr,\nonumber\\
    \Sigma^2&=(r^2+a^2)^2-2Mr,\nonumber\\
    \xi&=\frac{2 M a r }{\Sigma^2}.
\end{align}
We are interested in the singularity on equator plane $\cos\theta=0$, $r=0$, described in detail in \cite{Barrow:2019gzc}. For the explicit form of the extrinsic curvature scalars and Ricci scalar refer to \cite{Lee_2012}. Here, we only show the form of the Ricci scalar on the $\cos{\theta}=0$ plane:
\begin{align}
{}^{(3)}R=-\frac{2a^2 m^2(a^2+3r^2)^2}{r^4(r^3+a^2(2M+r))^2.}
\end{align}
It is singular at $r=0$. Integrating ${}^{(3)}R$ with the measure $N\sqrt{g}=r^2$, results in the infinite action in the UV limit and the Kerr spacetime does not contribute the path integral. The vanishing, 4-dimensional Ricci scalar is restored in the LI limit $\lambda=1$. It is then necessary to include the Kretschmann scalar to resolve the singularity as discussed in \cite{borissova2020blackhole}.
\paragraph{Dymnikova spacetime}
The Dymnikova spacetime is a regular solution in GR. It constructed with the line element (\ref{Heyward}) with:
\begin{align}
    f(r)=1-\frac{2M(r)}{r},\quad M(r)=M\left( 1-e^{-\frac{r^3}{2g^3}}\right).
\end{align}
The corresponding curvature scalars are non-singular:
\begin{align}
     {}^{(3)}R&=\frac{6M}{g^3} e^{-\frac{r^3}{2g^3}},\nonumber\\
    {}^{(3)}R_{ij}{}^{(3)}R^{ij} &=\frac{3 M^2}{2 g^6 r^6} e^{-\frac{r^3}{g^3}} \left(4 g^6 \left(e^{\frac{r^3}{2 g^3}}-1\right)^2\right.\nonumber\\
   &\left. -4 g^3 r^3 \left(e^{\frac{r^3}{2 g^3}}-1\right)+9 r^6\right)
\end{align}
and the action is finite in the limit $r_{UV}\xrightarrow{}0$. In particular, in this limit we have ${}^{(3)}R_{ij}{}^{(3)}R^{ij}\xrightarrow{}12 M^2/g^6$
\paragraph{Higher-order curvature scalars}
Here we give a general expression for the higher-order scalars present in the H-L potential for the projectable ADM and orthogonal gauge metric tensors.
The metric tensor in projectable ADM gauge (\ref{ADMgauge}) yields:
\begin{align}
    {}^{(3)}R_{ij}{}^{(3)}R^{ij}&=\frac{2 e^{-4 \nu (r)} \left(2 r^2 \nu '(r)^2+\left(r \nu '(r)+e^{2 \nu (r)}-1\right)^2\right)}{r^4},\nonumber\\
    {}^{(3)}R^i_{j}{}^{(3)}R^j_k {}^{(3)}R^k_i&=\frac{2 e^{-6 \nu (r)} \left(4 r^3 \nu '(r)^3+\left(r \nu '(r)+e^{2 \nu (r)}-1\right)^3\right)}{r^6}.
\end{align}
In the ortohonormal gauge they are given by
\begin{align}
{}^{(3)}R&=\frac{2 e^{-2 \Phi (r)} \left(2 r \Phi '(r)+e^{2 \Phi (r)}-1\right)}{r^2},\nonumber\\
    {}^{(3)}R_{ij}{}^{(3)}R^{ij}&=\frac{2 e^{-4 \Phi (r)} \left(2 r^2 \Phi '(r)^2+\left(r \Phi '(r)+e^{2 \Phi (r)}-1\right)^2\right)}{r^4},\nonumber\\
    {}^{(3)}R^i_{j} {}^{(3)}R^j_k {}^{(3)}R^k_i&=\frac{2 e^{-6 \Phi (r)} \left(4 r^3 \Phi '(r)^3+\left(r \Phi '(r)+e^{2 \Phi (r)}-1\right)^3\right)}{r^6}.
\end{align}
\subsection{Wormholes}
\renewcommand{\theequation}{B-\arabic{equation}}
\setcounter{equation}{0}
\paragraph{Einstein-Rosen bridge}
The Einstein-Rosen (E-R) bridge smoothly glues together two copies of the Schwarzschild spacetime: black-hole and the white-hole solutions corresponding to the 
positive and negative coordinate $u$.
Metric tensor of the Einstein-Rosen wormhole proposed in \cite{PhysRev.48.73} and discussed in e.g. \cite{Katanaev_2014} is given by:
\begin{align}
    ds^2=\frac{-u^2}{u^2+4M}dt^2+(u^2+4M)du^2+\frac{1}{4}(u^2+4M)d\Omega^2.
\end{align}
The E-R bridge is non-traversable and geodesically incomplete in $u=0$. This fact, however, does not impact the regularity of the curvature scalars.
The 4-dimensional Ricci scalar is:
\begin{align}
 R=\frac{2 \left(64 M^2+32 M u^2+4 u^4+u^2\right)}{\left(4 M+u^2\right)^3}.   
\end{align}
The second order curvature scalar $R_{\mu\nu}R^{\mu\nu}$ is:
\begin{align}
  \frac{4 \left(48 M^2+8 \left(4 M+u^2\right)^4+(32 M-1) \left(4 M+u^2\right)^2\right)}{\left(4 M+u^2\right)^6}.
\end{align}
Both of which integrated with the measure are non-singular:
\begin{align}
    \sqrt{g}=\frac{1}{4}u(4M+u^2).
\end{align}
The wormhole solutions analyzed in this paper generally yield the finite action in both GR and H-L. The finite Action Principle suggests, that in the quantum UV regime, singular black-hole spacetimes may be replaced with the regular wormhole solutions.
\paragraph{Morris-Thorne wormhole}
The MT wormhole is defined in the spherically symmetric, Lorentzian spacetime by the line element:
\begin{align}
    ds^2=-e^{2\Phi(r)}dt^2+\frac{dr^2}{1-\frac{b(r)}{r}}+r^2 d\Omega^2
\end{align}
where $\Phi(r)$ is known as the redshift and there are no horizons if it is finite. Function $b(r)$ determines the wormhole's shape. We choose $\Phi(r),\,b(r)$ to be:
\begin{align}
    \Phi(r)&=0,\quad
    b(r)=2M\left(1-e^{r_0-r}\right)+r_0 e^{r_0-r},
\end{align}
where $r_0$ is the radius of the throat of the wormhole, such that $b(r_0)=r_0$. 4-dimensional curvature scalars for this spacetime have been calculated in \cite{Mattingly_2020}. The Ricci curvature scalar is singular at $r=0$, however, the radial coordinate $r$ varies between $r_0>0$ and infinity:
\begin{align}
\label{RicciM-T}
    R=-2\left(2M-r_0\right)\frac{e^{r_0-r}}{r^2}.
\end{align}
The resulting $S_s=\int_{r_{UV}=r_0}^{r_{IR}} \sqrt{g}R$ function is divergent as $r_{UV}\xrightarrow{}r_0$ and cannot be expressed in terms of simple functions:
\begin{align}
    2(2M-r_0)\int_{r_{UV}=r_0}^{r_{IR}}\sqrt{\frac{r}{r-2M(1-e^{r_0-r})+r_0 e^{r_0-r}}}e^{r_0-r} dr.
\end{align}
However, this is only a coordinate singularity and one may get rid of it with a proper transformation.\\
Higher-order curvature scalars for Morris-Thorne wormhole are:
\begin{widetext}
\begin{align}
    {}^{(3)}R&=\frac{2 b'(r)}{r^2}\nonumber\\
    {}^{(3)}R_{ij}{}^{(3)}R^{ij}&=\frac{3 r^2 b'(r)^2-2 r b(r) b'(r)+3 b(r)^2}{2 r^6},\nonumber\\
    {}^{(3)}R^i_{j}{}^{(3)}R^j_k {}^{(3)}R^k_i&=\frac{-9 r^2 b(r) b'(r)^2+5 r^3 b'(r)^3+15 r b(r)^2 b'(r)-3 b(r)^3}{4 r^9},
\end{align}
\end{widetext}
and integrated give action that is finite.
\paragraph{H-L wormhole}
Static spherically traversable symmetric wormholes have been constructed in \cite{Botta_Cantcheff_2010} in the H-L theory through the modification of the Rosen-Einstein spacetime:
\begin{align}
    ds^2=-N^2(\rho)dt^2+\frac{1}{f(\rho)}d\rho^2+(r_0+\rho^2)^2 d\Omega^2,
\end{align}
with additional $\mathbf{Z}_2$ symmetry with respect to the wormhole's throat. There are solutions with $\lambda=1$ asymptotically corresponding to the Minkowski vacuum. Explicitly we have:
\begin{align}
    f=N^2&=1+\omega (r_0+\rho^2)^2\nonumber\\&-\sqrt{(r_0+\rho^2)\left(\omega^2(r_0+\rho^2)^3+4\omega M\right)}.
\end{align}
Radius of the wormhole's throat is given by $r_0$. The parameters $\omega,$ and $M$ are connected to the coupling constants in H-L action. See \cite{Botta_Cantcheff_2010} for their explicit form.
Ricci scalar of the H-L wormhole invariants are given by
\begin{widetext}
\begin{align}
    {}^{(3)}R&=-\frac{1}{{(\rho
   ^2+r_0){}^2}}\Big[2 (-10 \rho ^2 \sqrt{\omega  (\rho ^2+r_0) (4 M+\omega  (r_0+\omega ^2){}^3)}\nonumber\\&-4 r_0 \sqrt{\omega  (\rho ^2+r_0) (4
   M+\omega  (r_0+\omega ^2){}^3)}+16 \rho ^6 \omega +8 \rho ^2\nonumber\\&+36 \rho ^4 r_0 \omega +24 \rho ^2 r_0^2 \omega +4 r_0^3 \omega +4 r_0-1)\Big],\nonumber\\
   {}^{(3)}R_{ij}{}^{(3)}R^{ij}&= \frac{1}{(\rho ^2+r_0)^4}\left\{2 (-7 \rho ^2 \sqrt{\omega  (\rho ^2+r_0) (4 M+\omega  (r_0+\omega ^2){}^3)}-2 r_0 \sqrt{\omega  (\rho ^2+r_0) (4 M+\omega
    (r_0+\omega ^2){}^3)}\right. \nonumber\\& +10 \rho ^6 \omega 
    +6 \rho ^2+22 \rho ^4 r_0 \omega+14 \rho ^2 r_0^2 \omega +2 r_0^3 \omega +2 r_0-1){}^2+4 M \nonumber\\&
    +\frac{4 (\rho ^2+r_0)}{\omega  (4 M+\omega  (r_0+\omega ^2){}^3)}\left[\rho ^2 \omega  (-4 (\rho ^2+r_0) \sqrt{\omega  (\rho ^2+r_0) (4 M+\omega  (r_0+\omega ^2){}^3)}+\omega 
   (r_0+\omega ^2){}^3)\right]\nonumber\\&
   \left.-2 \sqrt{\omega  (\rho ^2+r_0) (4 M+\omega  (r_0+\omega ^2){}^3)} \left[-\sqrt{\omega  (\rho
   ^2+r_0) (4 M+\omega  (r_0+\omega ^2){}^3)}+\omega  (\rho ^2+r_0){}^2+1)\right]{}^2\right\}
\end{align}
\end{widetext}
The kinetic terms with $K_{ij}=0$ are vanishing, while the spacial Ricci scalar and higher curvature terms are finite.\\ From the point of view of the Finite Action Principle, all of the investigated wormhole spacetimes are included in the gravitational path integral. 

\addcontentsline{toc}{section}{The Bibliography}
\bibliography{Ree}{}
\bibliographystyle{apsrev4-1}
\end{document}